\documentclass{aastex}
\usepackage{spr-astr-addons}
\usepackage{graphicx}

\begin{document}

\title{A Predicted Relation between the Temperature and Density
Profile of Cluster Hot Gas}
\shorttitle{A Predicted Relation between Temperature and Density Profile}
\shortauthors{Chan and Chu}

\author{M.~H.~Chan} \and \author{M.-C.~Chu} 
\affil{Department of Physics and Institute of Theoretical Physics, The 
Chinese University of Hong Kong\\
Shatin, New Territories, Hong Kong, China}
\email{mhchan@phy.cuhk.edu.hk, mcchu@phy.cuhk.edu.hk}

\begin{abstract}
Based on the assumptions that a fraction of cluster dark matter is 
composed of degenerate neutrinos and they are in hydrostatic equilibrium 
with other matter, we predict a relation between the density profile and 
temperature of the cluster hot gas. The predicted relation agrees with 
observational data of 103 clusters.
\end{abstract}


\section{Introduction}
Observational data on rotation curves of galaxies and mass 
profiles of clusters indicate that dark matter exists. For example, the 
rotation curves of dwarf 
galaxies indicate that their total masses are much greater than the 
visible mass \citep{Swaters}, \citep{Salucci}. Also, the integrated total 
mass of a cluster is 
several times greater than the visible mass including hot gas and galaxies 
\citep{Reiprich}. On the other hand, recent neutrino oscillation 
experiments indicate that neutrinos have finite but small rest mass. 
Therefore, at least some 
fractions of dark matter should compose of neutrinos, which is 
known as hot dark matter (HDM). It is commonly believed that neutrinos 
exist in clusters and affect their structures. \citet{Cowsik} 
provided a simple model to understand the virial mass discrepancy in the 
Coma cluster if neutrinos have rest mass. After the non-zero rest mass of 
neutrinos was confirmed by experiments \citep{Fukuda,Bilenky}, neutrinos 
being a candidate cluster dark matter has become a hot topic again. 
\citet{Treumann} 
presented 
a model to calculate the mass distribution in two clusters, Coma and A119, 
including cold dark matter, $\sim$ 2 eV neutrino dark matter and hot gas.
Recently, \citet{Nakajima} presented 
a model using 1-2 eV degenerate neutrinos in hydrostatic equilibrium to 
fit the observed flat core in 
A1689, which contradicts with the results obtained by numerical 
simulations of cold dark matter particles \citep{Navarro,Moore}. 
All the above results indicate 
that neutrino dark matter can be an 
important component in the mass distribution of clusters. Neutrinos 
alone cannot form structures as their free streaming 
scale is too large. However, with the help of cold dark matter, neutrinos 
can be
gravitationally bound in the clusters and their effects may be 
observable \citep{formation}. 
In this article, we study a possible observable consequence of 
neutrinos in clusters. We assume 
that the degenerate neutrinos and hot gas particles are in hydrostatic 
equilibrium under 
the gravity of cold dark matter, galaxies and intergalactic hot gas. We 
derive an approximate relation 
among the parameters specifying the density profiles of the cluster hot 
gas and their temperatures in 103 clusters. We also make predictions 
about the density profiles of neutrinos in clusters. 

\section{Neutrinos in Clusters}
Currently, there are not much data on cluster observables. All we 
have 
now are the average hot gas temperature $T$, luminosity $L$, core radius 
$r_c$ and the parameter $\beta$ in King's $\beta$-model 
\citep{Brownstein}. In this section, we derive a relation among the 
cluster observables by 
assuming that neutrinos are bound in hydrostatic equilibrium by the 
overall mass profiles in clusters. 

In King's $\beta$-model \citep{King,Jones}, the hot gas number density is
\begin{equation}
n_g=n_c \left(1+ \frac{r^2}{r_c^2} \right)^{-3 \beta/2},
\end{equation}
where $n_c$ is the central number density. 
Suppose the hot gas is in hydrostatic equilibrium and the 
interactions between the baryons and the neutrinos are negligible; then 
the pressure gradients of the hot gas and the neutrinos are balanced by 
the total gravity inside a cluster independently. Therefore we have 
\begin{equation} 
\frac{kT}{m_gn_g} \frac{dn_g}{dr}=-\frac{GM(r)}{r^2},
\end{equation}
where $M(r)$ is the enclosed mass in a cluster including the masses of 
galaxies, hot gas, cold dark matter and neutrinos, $m_g$ is the average 
mass of the hot gas particles, and we have assumed that 
the temperature $T$ is constant throughout the 
hot gas. Although \citet{Vikhlinin} suggested the temperature may not be 
uniform especially near the center of some clusters, the variations may 
only amount to 20-30 \%, which has little effect on our results ($Tdn_g/dr 
\sim 4-5$ times greater than $n_gdT/dr$). Also, \citet{Chan} show that the 
temperature variation in cluster hot gas is not significant because
energy transfer by conduction is highly efficient. 
Therefore, in the following analysis, we follow \citet{Reiprich} and 
approximate the temperature as uniform. On the other hand, we 
suppose that the neutrinos with 
mass $m_{\nu}$ are degenerate and in hydrostatic equilibrium inside the 
cluster. Therefore we have
\begin{equation}
\frac{1}{\rho_{\nu}}\frac{dP}{dr}=-\frac{GM(r)}{r^2},
\end{equation}
where
\begin{equation}
P= \frac{4 \pi^2 \hbar^2}{5 m_{\nu}^{8/3}} \left( \frac{3}{4 \pi g_s} 
\right)^{2/3} \rho_{\nu}^{5/3}=K_{\nu} \rho_{\nu}^{5/3}
\end{equation}
being the degeneracy pressure of neutrinos, $\rho_{\nu}$ their mass 
density 
and 
$g_s$ the degree of freedom of each type of neutrinos. We assume 
$g_s=1$ and combine Eqs.~(3) and (4) to get
\begin{equation}
\frac{5K_{\nu} \rho_{\nu}^{-1/3}}{3}\frac{d \rho_{\nu}}{dr}=\frac{kT}{m_g} 
\frac{d(\ln{n_g})}{dr}.
\end{equation}
Using the density profile of the hot gas in Eq.~(1) and integrating
Eq.~(5), we finally obtain
\begin{equation}
\rho_{\nu}^{2/3}=\rho_c^{2/3}-\frac{3kT \beta}{5K_{\nu}m_g} 
\ln{\left(1+\frac{r^2}{r_c^2} \right)}
\end{equation}
for $r<R$ ($\rho_{\nu}=0$ for $r>R$) and the total enclosed mass profile 
\citep{Reiprich}
\begin{equation}
M(r)= \frac{3kTr^3 \beta}{m_gG(r_c^2+r^2)},
\end{equation}
where
\begin{equation}
\rho_c=\left[ \frac{3kT \beta}{5K_{\nu}m_g} \ln{\left(1+\frac{R^2}{r_c^2} 
\right)} \right]^{3/2}
\end{equation}
is the central neutrino density and $R$ is the 
radius of the neutrino density profile. The total mass 
profile Eq.~(7) has a soft core which is different from the NFW profile 
obtained by N-body simulation. Nevertheless, recent gravitational lensing 
data support the existence of soft cores in clusters, in
contradiction to the NFW profile 
\citep{Tyson,Sand,Broadhurst}. Since there is no robust definition of the 
radius and total 
mass of a cluster, we follow \citet{Brownstein} to define the radius and 
total mass of a cluster $M_c$ 
by assuming a cut off radius where the total mass density 
= 250 times mean cosmological density of baryons \citep{Brownstein}. We 
can then obtain 
a relation $\log M_{14}=(1.5 \pm 0.1) \log (\beta T_K)+ 
(-10.7 \pm 0.4)$ (see Fig.~1), where $M_{14}=M_c/10^{14}M_{\odot}$ and 
$T_K$ is the temperature of the hot gas in K, or
\begin{equation}
M_c \approx q( \beta T)^{3/2},
\end{equation}
where $q \approx (870-5010) M_{\odot}$ K$^{-3/2}$ is a constant which 
depends sensitively on the definition of the cut off radius.

\begin{figure}
\vskip5mm
 \resizebox{\hsize}{!}{\includegraphics[width=6cm]{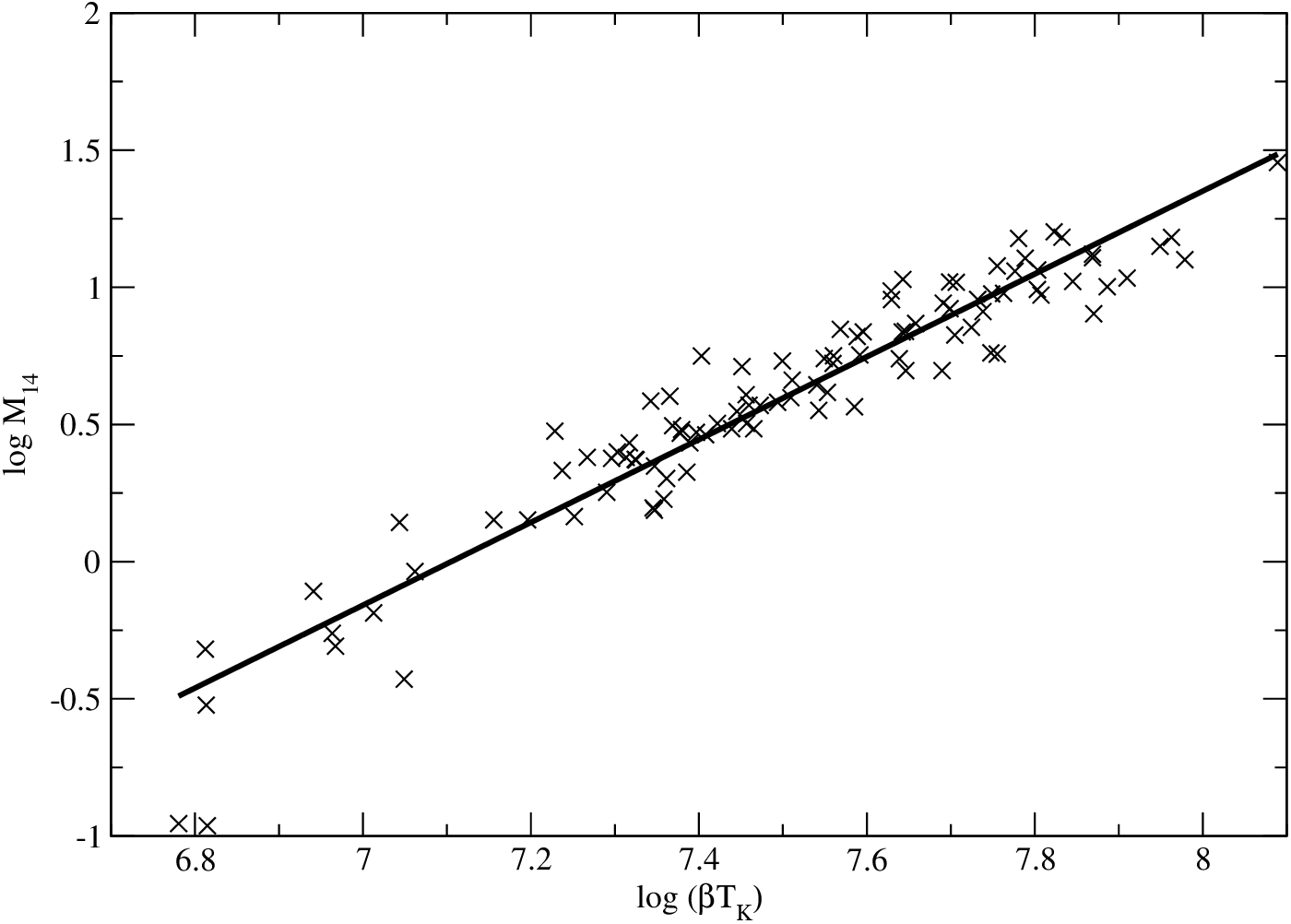}}
 \caption{$\log M_{14}$ against $\log (\beta T_K)$ for 103 
clusters. The solid line is the best fit line with a slope of $1.5 \pm 
0.1$.}
\end{figure}

In the following, we obtain a relation among the observables $r_c$, 
$\beta$ and $T$. We integrate the density profile in 
Eq.~(6) to get the total mass of the neutrinos:
\begin{equation}
M_{\nu}=
\int^R_04 \pi r^2 \rho_{\nu}dr
= \tilde{\rho}r_c^3I(u_0),
\end{equation}
where
\begin{equation}
I(u_0)= \int^{u_0}_04 \pi u^2 \left[\ln \left( \frac{1+u_0^2}{1+u^2} 
\right) \right]^{3/2}du,
\end{equation}
and $u=r/r_c$, $u_0=R/r_c$, $\tilde{\rho}=(3kT \beta 
/5K_{\nu}m_g)^{3/2}$.
In a cluster, we assume the ratio of $M_{\nu}$ to $M_c$ to be the same as 
the cosmological value
\begin{equation}
\frac{M_{\nu}}{M_c} 
\approx \frac{\Omega_{\nu}}{3 \Omega_m}= \frac{m_{\nu}}{\alpha},
\end{equation}
where $\alpha=94 \Omega_mh^2 \approx 12.6$ eV \citep{Chan}, $\Omega_{\nu}$ 
and 
$\Omega_m$ are cosmological density parameters of neutrinos and total 
matter, and $h \approx 0.7$ is the Hubble parameter.
By combining Eqs.~(8)-(12), we finally get:
\begin{equation}
r_c=\left[ \frac{qm_{\nu}}{\alpha I(u_0)} \right]^{1/3} \left( 
\frac{5K_{\nu}m_g}{3k} \right)^{1/2} \approx 
\frac{(2.0-3.6)~{\rm{Mpc}}}{m_a[I(u_0)]^{1/3}}.
\end{equation}
From Eq.~(13), we notice that for fixed $m_{\nu}$, $r_c$ depends on $u_0$ 
only. 
Plotting $R$ against $r_c$, we see that the values of $R$ 
are nearly constant for all clusters. $R$ is approximately 
proportional to $1/m_{\nu}$ (see 
Fig.~2). If $m_{\nu} \le 2$ eV, which is the current upper bound 
\citep{Oystein,Sanders}, then $R \gg r_c$ for most clusters. 
Suppose the total central density of a cluster $\rho_0$ is related 
to $\rho_c$ by a power law $\rho_c \sim \rho_0^{\gamma}$; by defining $4 
\pi r^2 \rho_0=dM(r)/dr$ at $r=0$ and rearranging Eq.~(8), we obtain
the key relationship between the core radius $r_c$ and the product $\beta 
T$:
\begin{equation}
\ln r_c \approx \left( \frac{2 \gamma -3}{4 \gamma} \right) \ln (\beta T) 
+{\rm constant},
\end{equation}
where we have assumed that $\ln \ln(1+R^2/r_c^2)$ is nearly a constant 
for all clusters. 
To verify the above prediction, we plot $\ln r_c$ against 
$\ln (\beta T)$ for 103 clusters in Fig.~3; an 
approximately linear relation is obtained which agrees with Eq.~(14). The 
slope in Fig.~3 is $0.97 \pm 0.11$ which corresponds to 
$\gamma \approx -3/2$ (correlation coefficient $=0.66$). However, the 
uncertainties in $M_c$, $\beta$, $T$ and 
$r_c$ are quite large, and the total mass profile of a cluster 
(Eq.~(7)) is only derived by using King's $\beta$-model. Therefore our 
model can only give an approximate prediction of the relation between $\ln 
r_c$ and $\ln (\beta T)$ with $\gamma \approx -3/2$.

\begin{figure}
\vskip5mm
 \resizebox{\hsize}{!}{\includegraphics[width=6cm]{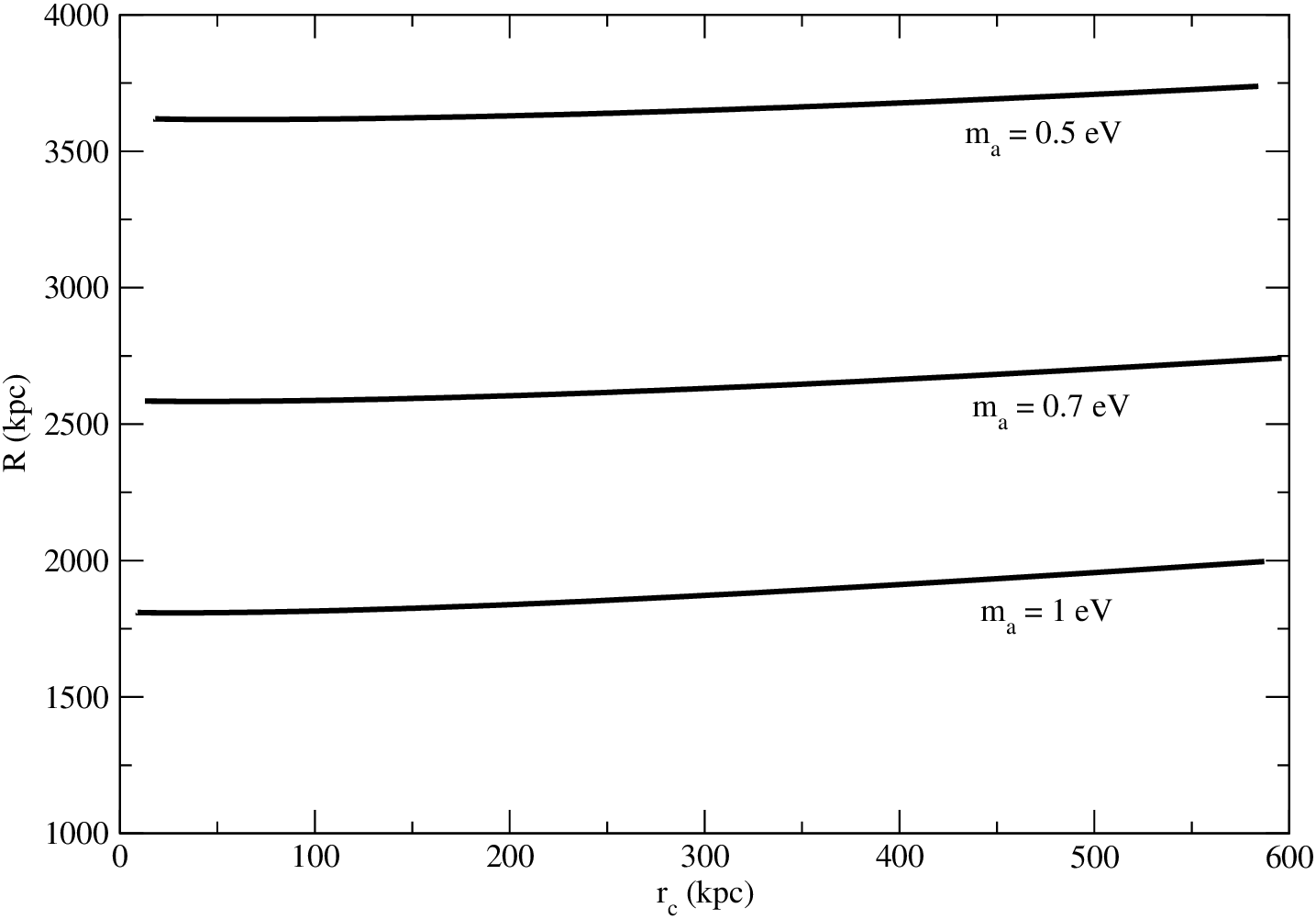}}
 \caption{$R$ against $r_c$ with $m_{\nu}$ = 0.5, 0.7 and 1 eV 
calculated by Eq.~(13).}
\end{figure}

\begin{figure}
\vskip5mm
 \resizebox{\hsize}{!}{\includegraphics[width=6cm]{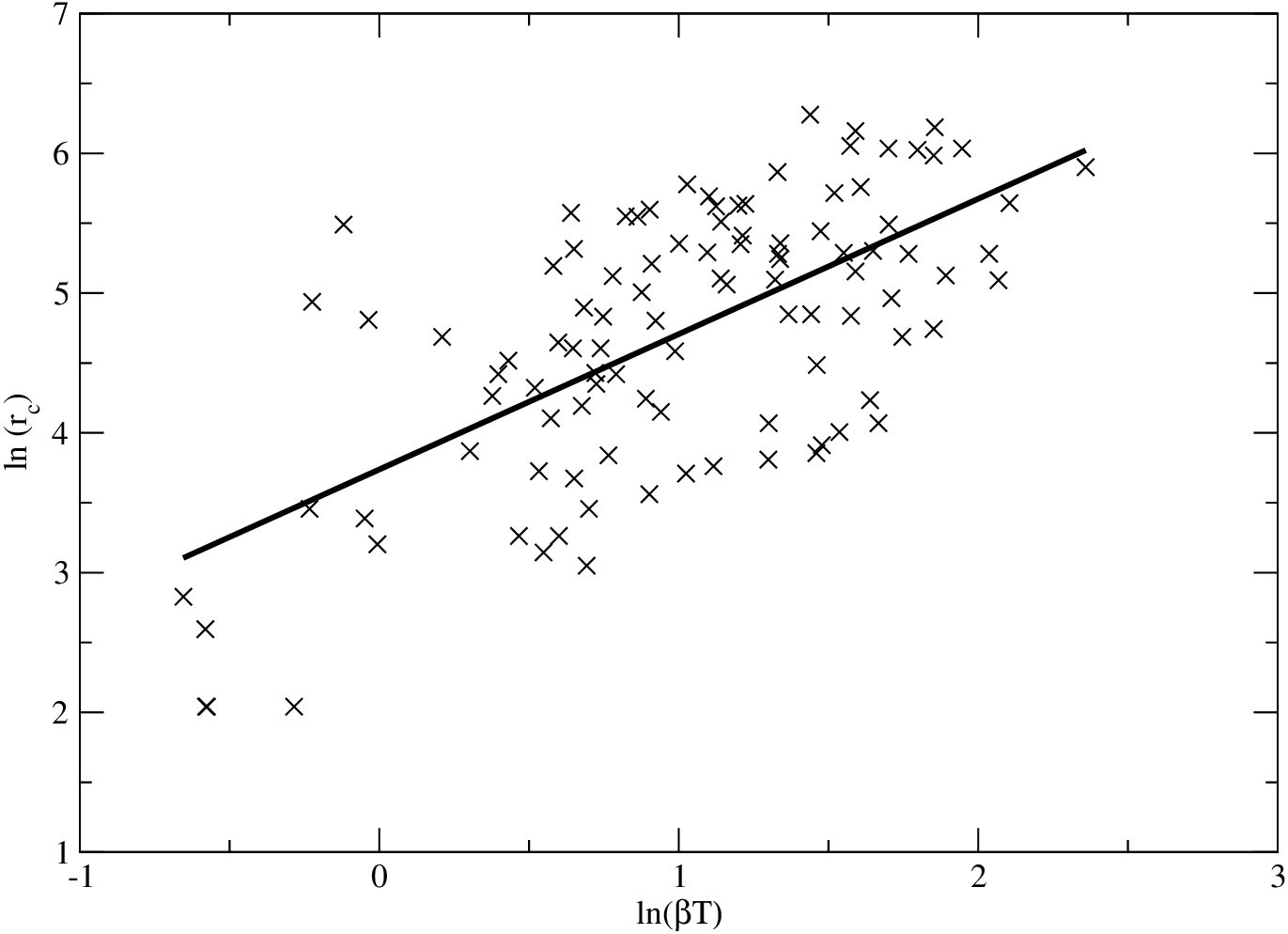}}
 \caption{$\ln r_c$ vs. $\ln (\beta T)$ for 103 
clusters, where $r_c$ and $T$ are in kpc and keV respectively. The crosses 
are the observed data and the solid line is the 
best fitted line. The slope obtained is $0.97 \pm 0.11$ with correlation 
coefficient $\sim 0.66$.}
\end{figure}

\section{Discussion and summary}
Neutrinos exist in clusters and they may form structures with help of cold 
dark matter \citep{formation}. By assuming the hydrostatic equilibrium of 
neutrinos and hot gas particles with total mass in 
clusters, we obtain the density profile of neutrinos in terms of $\beta$, 
$T$ and $r_c$, and we can thereby obtain an approximate 
relation among these parameters with $m_{\nu} \le 2$ eV. If $\rho_c 
\propto \rho_0^{\gamma}$, 
then a
linear relationship between $\ln r_c$ and $\ln \beta T$ is obtained which 
agrees with the observed data with $\gamma \approx -3/2$. Our result is 
also compatible with \citet{Sanders} that the core profiles in clusters 
can be explained by neutrinos as dark matter. 

\section{acknowledgements}
This work is partially supported by a grant from the Research Grant 
Council
of the Hong Kong Special Administrative Region, China (Project No. 
400805).

\bibliographystyle{spr-mp-nameyear-cnd}
\bibliography{biblio-u1}

\end{document}